\documentclass[conference]{IEEEtran}
\IEEEoverridecommandlockouts
\usepackage{cite}
\usepackage{amsmath,amssymb,amsfonts}
\usepackage{algorithmic}
\usepackage{graphicx}
\usepackage{textcomp}
\usepackage{xcolor}
\usepackage{svg}
\usepackage{multirow}
\def\BibTeX{{\rm B\kern-.05em{\sc i\kern-.025em b}\kern-.08em
    T\kern-.1667em\lower.7ex\hbox{E}\kern-.125emX}}
\begin{document}

\title{GRAFT: Graphlet-Triggered Backdoor Attack on GNN-Based Hardware Security Systems 
\thanks{This work was supported in part by the US National Science Foundation (NSF) Grant CNS-2228239. The views, opinions, and/or findings contained in this article are those of the author(s) and should not be interpreted as representing the official views or policies, either expressed or implied, of the US NSF.}
} 
\author{\IEEEauthorblockN{Sanaz Kazemi Abharian}
\IEEEauthorblockA{\textit{Department of Electrical and Computer Engineering} \\
\textit{George Mason University}\\
Fairfax, VA, USA \\
skazemia@gmu.edu}
\and
\IEEEauthorblockN{Sai Manoj Pudukotai Dinakarrao}
\IEEEauthorblockA{\textit{Department of Electrical and Computer Engineering} \\
\textit{George Mason University}\\
Fairfax, VA, USA \\
spudukot@gmu.edu}
} 
\maketitle

\vspace{-1in}
\begin{abstract}
The globalization of the integrated circuit (IC) supply chain increases the risk of security threats, such as hardware Trojans (HTs) and the theft of intellectual property (IP). Graph Neural Networks (GNNs), among the most powerful deep learning methods for processing graph-structured data, have been widely adopted to detect such threats. However, GNNs are susceptible to backdoor attacks that can maliciously manipulate output predictions toward an adversarial target. These attacks are not only difficult to detect but also compromise the integrity of GNN-based security systems. \textcolor{black}{Most prior work embeds backdoor triggers using randomly generated subgraphs or gradient-guided generative subgraphs. However, such triggers are impractical for GNN-based hardware security applications as they do not guarantee the preservation of circuit functionality. In this paper, we propose GRAFT, a graphlet-triggered backdoor attack targeting GNN-based hardware security. GRAFT embeds graphlet-based triggers at either the register-transfer level (RTL) or gate level of the design while preserving the circuit's original function.} We evaluate GRAFT on the ISCAS-85 and TrustHub datasets. Our experimental results demonstrate that GRAFT can effectively evade HT detection and IP piracy detection, achieving an attack success rate (ASR) of up to 100\%.
\end{abstract}


\section{Introduction}
The globalization of the integrated circuit (IC) supply chain has decentralized design, fabrication, and testing processes, resulting in their outsourcing to geographically dispersed third-party entities that may not be fully trusted \cite{akter2023survey}. This shift has heightened the risk of security threats, including hardware Trojans (HTs) and intellectual property (IP) piracy \cite{akter2023survey}. HT insertion compromises IC integrity through malicious modifications that can disrupt critical functionality or leak sensitive data. Conversely, IP piracy enables the unauthorized replication, modification, or redistribution of IP, resulting in substantial financial losses for legitimate stakeholders. Beyond economic consequences, IP theft undermines confidence in the supply chain and ultimately compromises the integrity and reliability of hardware ecosystems \cite{jiang2024hwsim}.

 Traditional HT detection methods often depend on golden reference models or extensive manual inspection, which can be infeasible or unreliable in modern, large-scale designs \cite{dong2020hardware}. Although watermarking and fingerprinting provide mechanisms to mitigate IP piracy, they are not robust defenses. These techniques incur non-trivial hardware overhead and remain vulnerable to sophisticated adversaries capable of removing, forging, or circumventing the embedded identifiers \cite{tauhid2023survey}. In contrast, Graph Neural Network (GNNs)-based techniques provide a robust and scalable alternative by capturing the structural and topological properties of ICs to learn discriminative patterns indicative of malicious behavior \cite{ saravanan2024accelerating}. GNNs eliminate the need for golden references, generalize to unseen threat variants, and can effectively detect subtle structural anomalies introduced by Trojans or IP theft, thereby emerging as a powerful tool for proactive hardware security analysis \cite{thangellamudi2026hardware, saravanan2025profuzz}. \textcolor{black}{GroVe \cite{waheed2024grove} uses the internal embedding space of the GNN (formed based on IC traits) as a fingerprint, enabling precise ownership verification by distinguishing pirated models from independently trained ones. In contrast, GNNFingers \cite{you2024gnnfingers} provides a black-box fingerprinting framework that constructs task-specific query graphs whose outputs uniquely identify the legitimate model, while PreGIP \cite{dai2025pregip} embeds watermarks during the pretraining stage of GNN encoders so that any downstream reuse of a stolen model reliably reveals the owner’s watermark.}

Despite the effectiveness of GNNs in hardware security applications, they remain vulnerable to various adversarial threats \cite{GBFA_ISQED_2026, abharian2026bitflip}. Among these, backdoor attacks pose a significant challenge, as they embed hidden triggers during training that can later be activated to manipulate the GNN's predictions without affecting its performance on normal inputs \cite{zhang2021backdoor}. The compromised model performs normally on benign inputs, but misclassifies any input containing a specific trigger pattern into the attacker's target label. Such attacks are typically realized by poisoning a small portion of the training data with trigger–label pairs, making them stealthy and difficult to detect during standard evaluation.

In prior works \cite{zhang2021backdoor,xu2021explainability}, backdoor triggers are injected randomly (e.g., by randomly constructing subgraphs) 
into benign graph for misclassification.  Although these methods are simple to implement, they generally lead to inconsistent and unstable attack performance. \textcolor{black}{Gradient-guided generative methods are introduced in \cite{xi2021graph}, which generate subgraphs based on the model's gradient information.
This type of trigger generation requires prior knowledge, such as access to the target structure and parameters.} \textcolor{black}{These types of triggers are explored only for specific domains such as biomedical, social, and networked data. } 


Such attacks on GNN models can be concerning in the context of hardware security, where an adversary could conceal the presence of a hardware Trojan or misclassify a pirated design as legitimate \cite{alrahis2023tt}, thereby bypassing GNN-based security analysis methods. Moreover, backdoors are designed to be imperceptible during standard validation, ensuring that the compromised model appears trustworthy while retaining hidden malicious behaviors. The covert and targeted nature of these attacks underscores the need to investigate GNN backdoor vulnerabilities to ensure the reliability of hardware security methodologies that rely on GNNs.
\textit{However, one of the challenges in creating such a backdoor attack on the GNN model(s) used for hardware security lies in the fact that any modifications made to the graph (IC) need to preserve the original circuit functionality with minimal footprint, making existing trigger-generation approaches impractical.
}

\textcolor{black}{In \cite{alrahis2023tt}, the backdoor triggers are implemented as cascade XOR inversion chains inserted into Boolean circuits. Although these structures maintain functional correctness and produce full toggle coverage, their deterministic and repetitive XOR cascade pattern creates an artificial subcircuit that does not naturally appear in real RTL designs. As a result, these triggers may introduce detectable structural footprints, for example long sequences of XOR gates with fixed constants or uniform bitwise inversions that specialized analysis tools or coverage based anomaly detectors can identify as suspicious.}

\textcolor{black}{In this paper, we propose GRAFT, a backdoor attack on GNN-based hardware security systems. GRAFT designs backdoor triggers in the form of \textit{graphlets}, which are small, connected induced subgraphs of a larger graph, and embeds them into RTL or gate-level designs. These triggers are later converted into logic blocks within the Boolean circuit while preserving circuit's functionality. We evaluate the effectiveness of GRAFT on GNNs used for HT detection and IP-piracy detection.}

\section{Backgrounds and Related Works}
\subsection{Graph Neural Networks}
In graph classification, a GNN is trained to predict a single label for an entire graph. Let $\mathcal{G} = \{(G_1, y_1), (G_2, y_2), \ldots, (G_N, y_N)\}$  be a dataset of $N$ labeled graphs, where each $G_i$ is a graph and $y_i \in \mathcal{Y} = \{c_1, c_2, \ldots, c_L\}$ is its label from $L$ possible classes. A GNN model parameterized by $\theta$, denoted $f_\theta(\cdot)$ is trained to learn a function $F : \mathcal{G} \rightarrow \mathcal{Y}$ that maps each graph to its corresponding class label.
\subsection{Graphlets for GNNs}
\begin{figure}[t]
      \centering
      \includegraphics[width=0.6\linewidth]{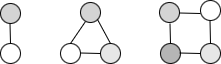}
      \caption{Illustration of graphlets with 2-4 node}
      \label{graphlet} 
\end{figure}

\textcolor{black}{Graphlets are small, connected, induced subgraphs that capture recurring structural patterns within a larger graph. As shown in Figure \ref{graphlet}, graphlets with 2, 3, and 4 nodes represent distinct connectivity patterns. Their occurrence counts provide compact and informative representations of both local and global network topology. Therefore, graphlets have been widely used in domains such as neuroscience and social network analysis. In the context of backdoor attacks on GNNs, graphlets can serve as meaningful trigger patterns because they encode important structural properties of graphs. Unlike random subgraphs, graphlet-based triggers are more closely aligned with graph topology and function, enabling more effective and stable backdoor attacks  \cite{hovcevar2014combinatorial}.}



\subsection{GNNs for IP Piracy Detection and HTs Detection}
GNN-based IP piracy detection models gate-level or RTL circuits as graphs, where nodes represent logic gates or modules and edges denote signal paths.
Using this representation, GNNs learn discriminative features to distinguish pirated circuits from legitimate ones. 
GNN4IP \cite{yasaei2021gnn4ip} leverages graph similarity to detect potential IP theft, while HWSim \cite{jiang2024hwsim} maps circuits into an embedding space using GNNs and measures similarity via cosine distance under a triplet-margin learning framework. GNN-based approaches enable Trojan detection without requiring prior knowledge of the IP or Trojan structure. By converting RTL or gate-level designs into graph representations, GNNs extract topological features and perform graph-level classification to determine the presence of HTs. Frameworks such as GNN4TJ \cite{yasaei2021gnn4tj} and related graph-based methods demonstrate that GNNs can effectively capture long-range dependencies introduced by Trojan triggers and payloads for accurate detection.

\begin{figure*}[h]
    \centering
    \includegraphics[width=0.8\linewidth]{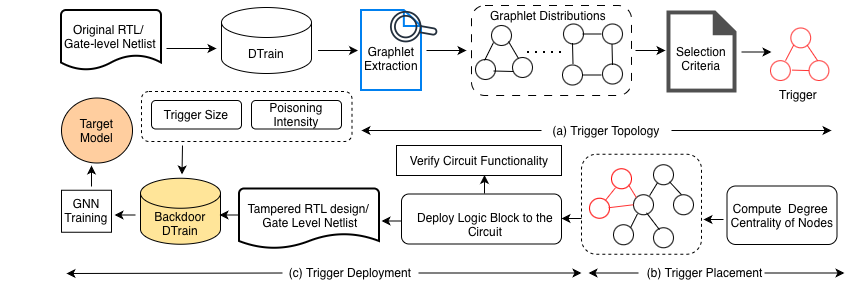}
    \caption{Overall framework of GRAFT. (a) Selection of the trigger topology based on graphlet statistics. (b) Selection of the trigger injection position based on node importance. (c) Logic block injection and GNN training using the backdoored dataset.}
    \label{GRAFT}
\end{figure*}
\begin{figure}[h]
    \centering
    \includegraphics[width=0.9\linewidth]{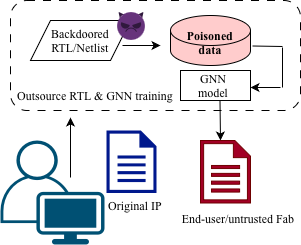}
    \caption{Threat model for injecting backdoor triggers. The adversary poisons RTL/ gate-level netlist.}
    \label{threat}
    \vspace{-0.2in}
\end{figure}
\section{GRAFT Attack Framework}
In this section, we present GRAFT, our proposed backdoor attack. Figure \ref{GRAFT} provides an overview of the GRAFT framework for injecting graphlet-based triggers into Boolean circuits. The framework consists of three main stages: trigger topology identification, trigger deployment, and trigger placement.
\subsection{Threat Model}
\textcolor{black}{ 
We consider an honest user (e.g., an IP vendor) sending its RTL/gate-level design, and the adversary gaining access to it. At this stage, the adversary can analyze the design in the form of Data Flow Graph (DFG) and determine where to insert triggers, as summarized in Figure \ref{threat}. Thus, the description, such as input size and number of layers, is sent to the trainer to train a GNN model, and the trainer returns the trained model. Since the user employs the GNN in a critical hardware security application (e.g., the IP piracy or HT detection), the user does not completely trust the prior stages. Thus, the user checks the performance of the trained GNN on a testing dataset. The user accepts the model if it meets the target accuracy value known as clean accuracy. The considered threat model is consistent with those commonly used in related research \cite{yang2024graph,gu2019badnets,wang2019neural}.} 


\subsection{Backdoor Trigger Topology} 


To design effective backdoor triggers, GRAFT first analyzes the structural representation of Boolean circuits. Each design is transformed into a DFG, where nodes correspond to logic gates and directed edges represent data dependencies between gates. This representation captures the functional flow of information in the circuit, while providing a graph structure for graphlet analysis.

GRAFT employs the Orbit Counting Algorithm (ORCA) to extract graphlet distributions from DFGs. ORCA algorithm counts the orbits that represent symmetry-equivalent positions within each graphlet \cite{hovcevar2014combinatorial}. This allows us to quantify how frequently different structural graphlets occur in the DFG and how individual nodes contribute to these graphlets. By examining orbit distributions, the adversary can identify underrepresented structures that are difficult for a graph to generalize over, as well as graphlets that exhibit bias toward particular labels. 
Then, GRAFT selects graphlet candidates that can serve as backdoor triggers. Thus, GRAFT denotes the process of choosing an appropriate graphlet topology. Based on this analysis, GRAFT prioritizes trigger topologies according to three criteria:
\begin{equation}
\tau = S(\mathcal{D}_{attack}),
\label{eq:trigger-selection}
\end{equation}
\par Where $D_{attack}$ is the dataset accessible to the adversary, and $S(\cdot)$ is the operator that selects the most suitable graphlet. This operator does not follow a single deterministic rule but instead applies a hierarchy of priorities:
\begin{enumerate}
 \item \textbf{Absence.} Graphlets that never occur in the dataset are selected first, since they represent out-of-distribution patterns that strongly influence model misclassification.
 \item \textbf{Sparsity.} Among candidate graphlets, those with fewer edges are favored, as they are easier to embed into circuits with minimal overhead and are less likely to be detected by functional verification.
 \item \textbf{Class bias.} Graphlets that appear disproportionately often in the target label are preferred, since their presence shifts the learned decision boundary toward the adversary's goal.
\end{enumerate}
\par In addition, the choice of graphlet size plays a central role in balancing stealth with attack effectiveness. Small graphlets such as triangles or three-node paths are lightweight and blend seamlessly into existing structures. Larger graphlets such as four-cycles or tailed triangles carry richer structural signatures that increase their influence on GNN predictions but risk greater functional perturbation if embedded carelessly. By restricting analysis to graphlets of size three to five, GRAFT ensures triggers are both expressive and concealable.

\begin{figure}[t]
    \centering
    \includegraphics[width= \linewidth]{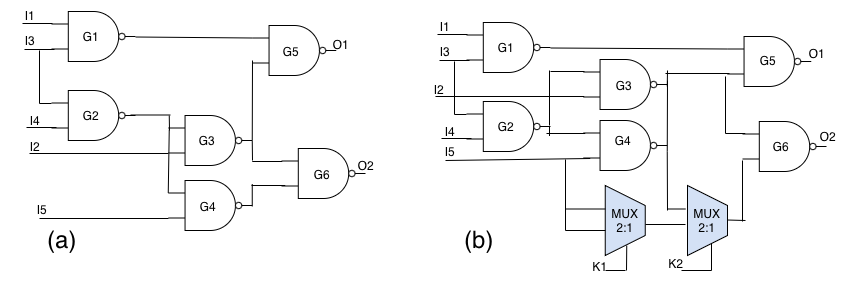} \vspace{-1em}
    \caption{Example of Logic Deployment in the Boolean circuit. (a) c17 from ISCAS-85 and (b) c17 with added logic blocks}
    \label{2} \vspace{-1.5em}
\end{figure}
\subsection{Backdoor Trigger Placement}
In addition to the structural design of the trigger, the position is a critical factor that significantly influences the effectiveness of a backdoor attack. After selecting the trigger topology, we further investigate how the choice of graphlet position influences the attack, considering both the graph structure and the characteristics of the GNN model. \textcolor{black}{To choose the position, the adversary does not access the GNN model. Thus, GRAFT employs a surrogate model to obtain target GNN model.}

GRAFT evaluates the \textcolor{black}{DFG} structure to select the most significant nodes in the graph as trigger placement candidates, as the graph structure encodes crucial information about the graph. Besides, GNNs operate based on message passing, relying on the graph structure for high performance. Thus, GRAFT uses the degree centrality (DC) parameter to measure the significance of nodes in the \textcolor{black}{DFG}, which is defined as:
\begin{equation}
    DC_i = \frac{d_i}{N - 1}
    \label{eq: DC}
\end{equation}
Here, $N$ is the number of nodes in the graph and $d_i$ denotes the degree of node $i$.

GRAFT sorts the nodes based on the highest DC index in descending order, then, selects $K$ top nodes as candidates for trigger \textcolor{black}{position.}
\paragraph{\textcolor{black}{Surrogate Model}} In addition to the graph structure, the significance of node information derived from the target model’s feedback should not be overlooked. The adversary often lacks access to the model’s internal details, such as its architecture and parameters. Still, the responses provided by the target model can effectively guide the attacker in crafting successful backdoored graphs. The outputs of the target model are essential for guiding the generation of backdoored graphs. 
To enable this limitation, GRAFT adopts a \textcolor{black}{surrogate} model, where a shadow model is trained to approximate the behavior of the \textcolor{black}{target GNN model} by utilizing its accessible predictions. Through this process, the \textcolor{black}{surrogate model} $F_{\theta}$ serves as a surrogate that distills the decision boundary and confidence distribution of the \textcolor{black}{target model}. The distilled knowledge enables the attacker to obtain reliable feedback for optimizing the backdoor trigger, thereby enhancing attack transferability to the target model.

\paragraph{Screening via \textcolor{black}{Surrogate} Model} Beyond structural placement, node selection can also be refined by leveraging the surrogate model (considered as a shadow model) to assess node significance. GRAFT employs a node-dropping strategy: each candidate trigger node is individually removed from the benign graph to construct a candidate subgraph set $G_{can}$. The surrogate model $F_{\theta}$ is then used to evaluate the difference in predictions between the original benign graph and each modified subgraph. This difference defines the node’s importance score (subscore). A higher subscore implies that the removed node plays a more critical role in the model’s decision process. Formally, the subscore is defined as:
\begin{equation}
\text{subscore}_{r} = \big|F_{\theta}(G + r) - F_{\theta}(G)\big|, \quad r \in N_{can},
\end{equation}
where $G$ is the benign graph, $r$ denotes the operation of dropping the $r$-th node, $F_{\theta}(\cdot)$ is the student (shadow) model, and $N_{can}$ is the set of candidate trigger nodes. 
From the model’s perspective, removing highly influential nodes leads to greater perturbations in output. 
Therefore, GRAFT ranks candidate nodes according to their subscores, ensuring that triggers are placed on nodes that exert maximal impact on model predictions.

\subsection{\textcolor{black}{Boolean Trigger for GNN Backdoor}} 
In this section, we explain in detail how to \textcolor{black}{select a proper Boolean circuit that acts as a backdoor trigger in GNN models} and launch an attack on the Boolean circuit. We integrate cascade multiplexers to launch the trigger backdoor attack on a Boolean circuit, which is translated to graphlets as subgraphs. For example, Figure \ref{2} illustrates the use of cascade multiplexers in the C17 circuit, a benchmark from ISCAS-85. Since we have two multiplexers with identical inputs, the outputs of the multiplexers do not change the design's functionality. However, it modifies the overall graph structure without affecting the design's functionality. 
Besides, we consider the steps for trigger placement and trigger topology to deploy logic blocks. We select a multiplexer to ensure that the functionality of the circuit does not change. Therefore, the multiplexer-based trigger is able to pass the test because it passes circuit-level functional and design rule checks, as the hardware behaves normally, and it can pass GNN-based security detection because the poisoned model misclassifies malicious circuits as benign.

We select the number of logic blocks as trigger based on the poisoning ratio, $p$, a corresponding number of backdoored graphs are generated and incorporated into the training dataset. These graphs are then involved in the target model’s training process, thereby influencing its parameters and embedding a backdoor, ultimately resulting in a compromised (backdoored) GNN model that will be used for hardware security analysis. 

\subsection{Time Complexity Assessment of GRAFT}
GRAFT 's time complexity incudes four main steps: DFG construction, graphlet-based trigger selection, trigger placement, and trigger deployment. For a dataset with $G$ graphs, where $G_i$ has $n_i$ nodes, it is computed according to Eq. \ref{eq:time}.
\begin{align}
T_{\text{GRAFT}} = \;&
\sum_{i=1}^{G} \mathcal{O}(n_i + m_i)
+ \sum_{i=1}^{G} \mathcal{O}(m_i \Delta_i^{k-2}) \nonumber \\
&+ \sum_{i=1}^{G} \big( \mathcal{O}(m_i) + \mathcal{O}(n_i \log n_i) \big)
+ \mathcal{O}(p N_{\text{train}}) .
\label{eq:time}
\end{align}

Here, $\mathcal{O}(n_i + m_i)$ accounts for DFG construction, $\mathcal{O}(m_i \Delta_i^{k-2})$ captures graphlet analysis for trigger selection, $\big( \mathcal{O}(m_i) + \mathcal{O}(n_i \log n_i) \big)$ represents degree-centrality computation and node ranking, and $\mathcal{O}\left(pN_{\text{train}}\right)$ denotes trigger deployment on the poisoned training samples.

\begin{table}[!t]
\centering
\caption{Statistics of the IP Piracy Detection Dataset}
\label{tab:gnn4ip-stats}
\setlength{\tabcolsep}{3pt}
\renewcommand{\arraystretch}{1.15}
\resizebox{\columnwidth}{!}{%
\begin{tabular}{|c|c|c|c|c|}
\hline
\textit{Baseline Designs} & \textit{Obfuscated Instances} &
\textit{Total \#Pairs} & \textit{\#Similar Pairs} & \textit{\#Different Pairs} \\
\hline
c432 & 23 & \multirow{3}{*}{2{,}701} & 
\multirow{3}{*}{890} & \multirow{3}{*}{1{,}811} \\
\cline{1-2}
c499 & 22 &  &  &  \\
\cline{1-2}
c880 & 29 &  &  &  \\
\hline
\end{tabular}%
}
\label{IPPiracy}
\end{table}


\begin{figure*}[h]
    \centering
    \includegraphics[width=0.8\linewidth]{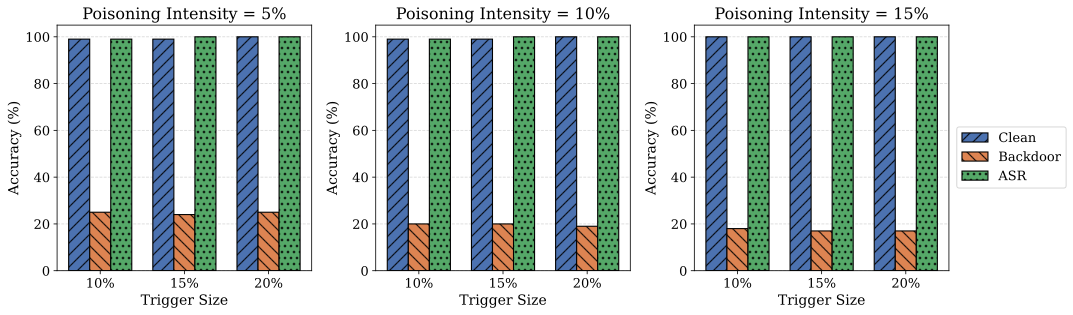} \vspace{-1em}
    \caption{Effect of Trigger size $(\phi)$ and Poisoning Intensity $(\gamma)$ on the Performance of GRAFT in GNN-Based IP Piracy Detection} 
    \label{6} \vspace{-1em}
\end{figure*}

\begin{figure*}[h]
    \centering
    \includegraphics[width=0.8\textwidth]{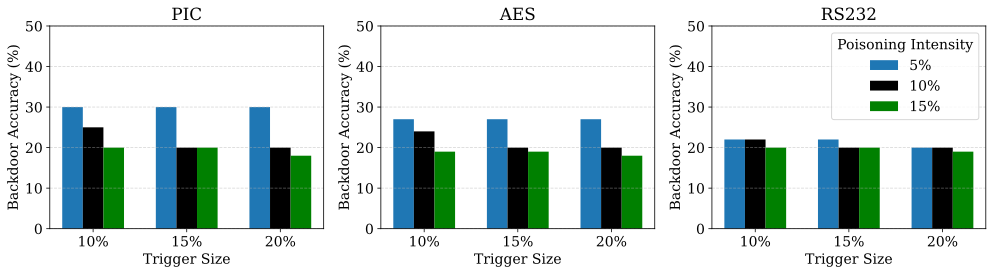}
    \caption{Impact of GRAFT under Varying Trigger Size $(\phi)$ and Poisoning Intensity $(\gamma)$ on GNN-Based HT Detection for PIC, AES,  and RS232 Datasets }\vspace{-1em}
    \label{4} 
\end{figure*}

\vspace{-1em}
\section{Experimental Evaluation and Analysis}
\subsection{Evaluation Metrics}
We evaluate the GRAFT using two metrics: accuracy and ASR. Clean accuracy measures performance on benign data, while backdoor accuracy measures performance when triggers are present. ASR denotes the probability that a compromised input is classified as the target class and is computed as:
\begin{equation}
\text{ASR} = \frac{N_{\text{suc}}}{N_{\text{att}}}
\end{equation}
\par Here, $N_{suc}$ indicates the number of samples that were successfully misclassified into the target class, and $N_{att}$ represents the total number of samples subjected to the attack.
\subsection{Experimental Settings}
We employ the TrustHub \cite{amir2018development} and ISCAS-85 benchmarks as datasets.  To ensure statistical reliability, we evaluate the performance of GRAFT and the baseline models over five independent runs and report the average results. \textcolor{black}{Backdoor trigger size $(S)$ refers the number of nodes in the backdoor trigger or subgraph. As various circuits have different graph size, for each circuit, the backdoor trigger size $(S)$ is defined as a fraction $\phi$ of the total number of nodes in the circuit. Moreover, poisoning intensity $\gamma$ indicates the percentage of training graphs that are poisoned by the adversary.}


\subsection{Graphlet Distribution Analysis}
We obtain the DFG representation of each Boolean circuit from its RTL or gate-level description using the Pyverilog parser. By applying the ORCA algorithm and counting the orbits of individual nodes, we derived the corresponding graphlet distributions across the dataset.
\subsection{Case Study 1: IP Piracy Detection}

We employ GNN4IP \cite{yasaei2021gnn4ip} to train a GCN model to detect IP piracy. This model takes two designs $(p_1, p_2)$ at the same time. Additionally, we use GCN settings of GNN4IP; 2 GCN layers with 128 hidden units. The parameters of the GCN are trained using a loss function $L$ according to the equation \eqref{3} defined on the training dataset $D$, which incorporates the cosine similarity between the designs.
\begin{equation}
L(\hat{y}, y) =
\begin{cases}
1 - \hat{y}, & \text{if } y = 1, \\
\max(0, \hat{y} - 0.5), & \text{if } y = -1 \, .
\end{cases}
\label{3}
\end{equation}
\par \textbf{Dataset:} Following GNN4IP, we use its original dataset, which contains three ISCAS-85 benchmarks (c432, c499, and c880) along with multiple hardware-obfuscated variants of each, yielding 74 gate-level netlists in total. Table \ref{IPPiracy} summarizes the characteristics of dataset employed for GNN to detect IP Piracy.

\textcolor{black}{Figure \ref{6} illustrates the impact of trigger size $(\phi)$ and poisoning intensity $(\gamma)$ on the performance of GRAFT in GNN-based IP piracy detection. The backdoor accuracy remains low, ranging approximately between 17\% and 25\% which confirms that the model misclassifies triggered samples as intended by the attack. Meanwhile, ASR consistently stays close to 100\% across all configurations, demonstrating the effectiveness of the injected trigger.}

\textcolor{black}{Increasing the trigger size from 10\% to 20\% has minimal impact on accuracy or ASR, indicating that the attack remains effective even with relatively small graphlet trigger patterns. However, the backdoor accuracy slightly decreases as poisoning intensity increases from 5\% to 15\% indicating that higher poisoning levels strenghten the attack by reducing the model 's ability to correctly classify triggered inputs.}




\subsection{Case Study 2: Hardware Trojan Detection}
We use GNN4TJ framework \cite{yasaei2021gnn4tj} to train a two-layer GCN with 200 hidden units each. The model uses the generated embedding for the graph, to make a prediction $\hat{y}$, either TjIn or TjFree via minimizing the cross entropy loss $L$ for all the graphs in $D_{Train}$.
\begin{equation}
L(\{y_G\}, \{\hat{y}_G\}) = \sum_{G} y_G \cdot \log_{e}(\hat{y}_G) \, .
\end{equation}
\par \textbf{Dataset:} We examine the GRAFT in hiding HTs on three datasets: RS232, PIC microcontroller, and AES from TrustHub \textcolor{black}{and HT-free samples such as DET, RC6, and XTEA circuits are added to balance the dataset.} A summary of the datasets' statistics is provided in Table \ref{HT}.

\textcolor{black}{To evaluate the effectiveness of GRAFT for detecting HTs using GCN, we inject different values of varies $\phi$ and $\gamma$. As shown in Figure \ref{4}, even with a poisoning intensity as low as 5\%, the backdoor accuracy drops to approximately 30\% across all trigger sizes. Moreover, the ASR remains consistently high (98-100\%) across all trigger sizes and poisoning intensities.}


\paragraph*{Proof of preserving Boolean circuit functionality} We verified the functionality of the modified netlist after adding the backdoor triggers- for the training dataset, using $100K$ input vectors and the functionality did not change and we could achieve 100 \% functionality preservation.
\begin{table}[ht]
\centering
\vspace{-1em}
\scriptsize 
\setlength{\tabcolsep}{3pt}
\caption{Statistics of the HT Detection Dataset.}\vspace{-1em}
\begin{tabular}{|c|c|c|c|c|}
\hline
\textbf{Dataset} & \textbf{\#Classes} & \textbf{\#Graphs} &
\textbf{\#Nodes in base circuit} & \textbf{\#Testing graphs} \\
\hline
PIC   & 2 & 29 & 2541  & 5 \\
\hline
RS232 & 2 & 29 & 668   & 8 \\
\hline
AES   & 2 & 29 & 14007 & 5 \\
\hline
\end{tabular}
\label{HT}\vspace{-2em}
\end{table}
\subsection{Comparison with State-of-the-Art Methods}
\textcolor{black}{We compare the GRAFT with PoisonedGNN \cite{alrahis2023tt}, the only prior work to ours. Both approaches implement backdoor triggers as sub-circuits in GNN-based hardware security systems. However, they employ different trigger mechanisms and circuit structures. In PoisonedGNN, triggers are inserted at predefined locations in the output gate as cascaded XOR gates. However, this approach increases the likelihood of detection during hardware testing. In contrast, multiplexer-based triggers provide better stealth and structural flexibility, as they resemble legitimate hardware logic and are therefore less suspicious.}

\section{Conclusion}
\textcolor{black}{This paper introduces GRAFT, a backdoor attack on GNN-based hardware security systems that injects graphlet-based triggers into RTL or gate-level circuit representations while preserving circuit functionality. These backdoor triggers are translated into sub-circuits within the corresponding Boolean circuits. GRAFT achieves a 100\% ASR by misclassifying HT-infected and pirated circuits as benign.}
\vspace{-0.5em}
\bibliographystyle{IEEEtran}
\bibliography{reference}
\vspace{12pt}
\end{document}